\documentclass[12pt]{article}
\usepackage{hyperref}
\usepackage{amsmath}
\usepackage{graphicx}
\usepackage{amssymb} 
\newcommand{\Mat}[1]{{{\boldsymbol{#1}}}}
\newcommand{\abs}[1]{\left\vert#1\right\vert}
\def\be{\begin{equation}}
\def\ee{\end{equation}}
\def\bea{\begin{eqnarray}}
\def\eea{\end{eqnarray}}
\def\dd{\mathrm{d}}
\title{Dirac equation: Representation independence and tensor transformation}

\author{
Mayeul Arminjon\,$^1$ and Frank Reifler\,$^2$\\
$^1$ \small\it Laboratoire ``Sols, Solides, Structures, Risques'' (CNRS \& Universit\'es de Grenoble),\\
\small\it BP 53, F-38041 Grenoble cedex 9, France.\\
$^2$ \small\it Lockheed Martin Corporation, MS2 137-205,\\ 
\small\it 199 Borton Landing Road, Moorestown, New Jersey 08057, USA.
} 
\date{ }

\begin{document}

\maketitle

\begin{abstract}
We define and study the probability current and the Hamiltonian operator for a fully general set of Dirac matrices in a flat spacetime with affine coordinates, by using the Bargmann-Pauli hermitizing matrix. We find that with some weak conditions on the affine coordinates, the current, as well as the spectrum of the Dirac Hamiltonian, thus all of quantum mechanics, are independent of that set. These results allow us to show that the tensor Dirac theory, which transforms the wave function as a spacetime vector and the set of Dirac matrices as a third-order affine tensor, is physically equivalent to the genuine Dirac theory, based on the spinor transformation. The tensor Dirac equation extends immediately to general coordinate systems, thus to non-inertial (e.g. rotating) coordinate systems.
\\


\end{abstract}

\section {Introduction}

The Dirac equation is associated with a specific transformation behaviour: although it has four components, the Dirac wave function $\psi $ is not transformed as a spacetime vector. Instead, it is subjected to the so-called spinor transformation: $\psi \mapsto S\psi $, which arises because the Dirac matrices $\gamma ^\mu$ are supposed to stay invariant after a Lorentz transformation $L$. The spinor representation: $L \mapsto S={\sf S}(L)$, is defined (unambiguously up to a sign) for $L \in {\sf SO}^+(1,3)$, the proper orthochronous Lorentz group, but it cannot be extended to the group of general linear transformations, ${\sf GL}(4,{\sf R})$ \cite{Weyl1929b, Scholz2004, Doubrovine1982}. This means that the use of the genuine Dirac equation is limited to Cartesian coordinates. Thus, for instance, the genuine Dirac equation cannot be used to describe the situation in a rotating frame, which is relevant to Earth-based experiments. In such non-inertial frames, one has to use \cite{Mashhoon1988,HehlNi1990,VarjuRyder2000,Obukhov2001,A41} the extension of the Dirac equation proposed independently by Weyl \cite{Weyl1929b} and by Fock \cite{Fock1929b}, hereafter the ``Dirac-Fock-Weyl" (DFW) equation. However, the DFW equation does not transform as the Dirac equation under a coordinate change: for the DFW equation, the wave function $\psi $ stays {\it invariant} after any coordinate change, while the (ordered) set $(\gamma ^\mu)$ transforms as a mixed object which is only partially tensorial \cite{BrillWheeler1957+Corr, deOliveiraTiomno1962}. Now one lesson of relativity is that the physical consequences of an equation may depend, not only on the equation itself, but also on its transformation behaviour---for instance, the Maxwell equations do not describe the same physics, depending on whether Galileo transformation or Lorentz transformation is used. Therefore, it is not a priori obvious that, if one neglects gravitation (thus assuming a flat spacetime), the DFW equation is physically equivalent to the Dirac equation.\\

It turns out to be possible \cite{A37} to transform the {\it usual} Dirac equation {\it covariantly,} with the wave function transforming as a {\it spacetime vector,} i.e., 
\be \label{psi-vector}
\psi '=L\psi \quad (\psi '^\mu = L^\mu_{\ \nu}\, \psi ^\nu), \qquad L^\mu_{\ \nu} \equiv \frac{\partial x'^\mu }{\partial x^\nu },
\ee
provided one simultaneously transforms the Dirac matrices $\gamma ^\mu $ in the following way:
\be \label{gamma-(^2_1)tensor}
\gamma '^\mu \equiv L^\mu_{\ \nu}\, L \gamma ^\nu L^{-1}.
\ee
If, as usual, one writes the row index of the Dirac matrices as a superscript and the column index as a subscript, this equation means simply that the threefold array of the components, $\gamma^{\mu \rho} _\nu\equiv \left(\gamma^\mu \right)^\rho _{\ \nu}$, is a $\ {\it (^2 _1)}$ {\it tensor} \cite{A39}. The very possibility of applying a tensorial transformation to the usual Dirac equation itself had not been recognized before---even though there have been attempts at {\it rewriting} the Dirac equation in a different form and with different fields, so as to recover tensors: e.g. Eddington \cite{Eddington1928,Durham2005}, Whittaker \cite{Whittaker1937} (see also Taub \cite{Taub1939}), Elton \& Vassiliev \cite{Elton-Vassiliev1998}. In a presentation of the Dirac equation starting from a special choice for the $\gamma ^\mu $ matrices (involving four ``Pauli-like" $2\times 2$ matrices $\sigma ^\mu $), Bade \& Jehle \cite{BadeJehle1953} envisaged a (peculiar) possibility in order that the Dirac equation be covariant with respect to Lorentz transformations with non-fixed matrices. But they immediately dismissed it, on the ground that ``the $\sigma ^\mu $ will have the especially simple [Pauli-like] values only in certain frames of references." This argument might be put forward in the same way against the general tensor transformation (\ref{psi-vector})-(\ref{gamma-(^2_1)tensor}): once the $\gamma ^\mu $ matrices are changed after a coordinate change, a special choice for them can be taken only in special coordinate systems---which would seem to ``violate the spirit if not the letter of the relativity idea" \cite{BadeJehle1953}.\\

However, another important point is that the choice of the $\gamma ^\mu $ matrices should not have any physical consequence, provided they fulfil the relevant anticommutation relation,
\be \label{Clifford}
\gamma ^\mu \gamma ^\nu + \gamma ^\nu \gamma ^\mu = 2g^{\mu \nu}\,{\bf 1}_4, \quad \mu ,\nu \in \{0,...,3\}.
\ee
As it appears more clearly in some derivations of the Dirac equation, which adopt the 4-dimensional covariant formalism from the beginning, there is indeed no reason to prefer any set of anticommuting matrices (see e.g. Refs. \cite{A37,Schulten1999}). If it turned out that the physical predictions of the Dirac equation should depend on the set $(\gamma ^\mu)$, this would certainly invalidate the transformation behaviour (\ref{psi-vector})-(\ref{gamma-(^2_1)tensor}), but it would be also a very serious problem for the standard (spinor) transformation: one would have to find good reasons to select a special choice, say $(\gamma ^\mu_0)$, and this would have to be made once and for all. It is surprising that the possible dependence of the physical predictions on the set $(\gamma ^\mu)$ seems to be hardly discussed in the literature: almost always, some particular choice is made, generally the ``standard" choice, e.g. Bjorken \& Drell \cite{BjorkenDrell1964}. Even when some other sets are presented, such as the so-called ``chiral" matrices (e.g. Schulten \cite{Schulten1999}) or the Majorana matrices \cite{MohapatraPal2004}, no attempt is made at showing that the physical predictions are unaffected by this or another choice. Recently, Pal \cite{Pal2007} has derived various important identities involving Dirac matrices and spinors, independently of any particular choice $(\gamma ^\mu)$, which represents a step towards showing the ``representation independence." However, he restricts the consideration to sets such that $\alpha ^0\equiv \gamma ^0$ and $\alpha^j\equiv \gamma ^0\gamma ^j\ (j=1,2,3)$ are Hermitian {\it matrices,} i.e., $\alpha ^{\mu\, *T}=\alpha ^\mu$. This forces him to consider similarity transformations that are defined by a unitary matrix, whereas this limitation is not imposed by the anticommutation (\ref{Clifford}), nor---as will be shown here---by the condition that the Hamiltonian be a Hermitian operator (this is because the scalar product has to be specified). Moreover, Pal's results \cite{Pal2007} do not directly enable one to answer the following questions: 
\begin{itemize}
\item Does the probability current depend on the chosen set $(\gamma ^\mu)$?

\item Is there a positive definite inner product defined for the wave functions relevant to the Dirac equation ?
  
\item Does the inner product depend on the chosen set $(\gamma ^\mu)$?

\item Does the spectrum of the Hamiltonian operator depend on the chosen set $(\gamma ^\mu)$?

\end{itemize}
In our opinion, these are crucial questions. The first part of the present paper (Sects. \ref{current} to \ref{uniqueness}) will be devoted to answering them favorably, with some minor, but important, restrictions on the affine coordinates discussed in Section \ref{Hamiltonian}.  To do this, we present all the elements of quantum mechanics, including the Dirac equation, positive probability density and its conserved current, positive scalar product for a Hilbert space, and the Hamiltonian, which we show is a Hermitian operator, in affine coordinate systems that are not Cartesian.  This will enable us then to show, after a summary of previous work on the tensor transformation of the Dirac equation \cite{A37} (Sect. \ref{transformation-options}), that the latter is physically completely equivalent to the spinor transformation as long as only Lorentz transformations are allowed (Sect. \ref{linear-change})---with the advantage that the tensor transformation extends to the case of affine coordinate transformations (Sect. \ref{linear-change}) and to the case of general coordinate changes (Sects. \ref{linear-change} and \ref{general-change}).\\

In the appendices we prove some new extensions of Pauli's Theorems \cite{Pauli1933,Pauli1936}, which have not been considered by previous authors, but are needed to extend the Hilbert space for the Dirac equation to affine coordinates, and to get the positive definiteness and uniqueness of the inner product.  Part of these results, with the notable exception of the positive definiteness of the inner product, might be derived more or less directly from the works of Pauli \cite{Pauli1933,Pauli1936} and Kofink \cite{Kofink1949}, though not in the general case that we need. We show that the positive definiteness of the inner product is not valid in general, but requires some weak conditions on the affine coordinates.  These weak conditions are always satisfied in practice in admissible spacetimes (see Sect. \ref{Hamiltonian}).
\footnote{\ 
Theorem 4 and part of Theorem 5 that depends on Theorem 4 in Appendix \ref{uniqueness-proof} are new.  Theorems 6  and 7 in Appendices \ref{uniqueness-affine} and \ref{anticom-alpha} are also new. 
}

\section{Current conservation for a general set of Dirac matrices}\label{current}

Let us consider a general set $(\gamma ^\mu)$ of Dirac matrices, i.e., an ordered set of four $4 \times 4$ complex matrices satisfying the anticommutation relation (\ref{Clifford}), with $(g^{\mu \nu})\equiv(g_{\mu \nu})^{-1}$, where $(g_{\mu \nu})$ is the matrix of the components of a general metric $\Mat{g}$, that is, a non-degenerate, real, and symmetric spacetime tensor. (Spacetime indices will be lowered and raised using the metric $g_{\mu \nu}$ and its inverse $g^{\mu \nu}$.) The main tool to deal with a such general set is the {\it hermitizing matrix,} first introduced for a particular set by Bargmann \cite{Bargmann1932}, and studied in a more general case by Pauli \cite{Pauli1933,Pauli1936}. This is a nonzero $4 \times 4$ complex matrix $A$ such that
\be\label{hermitizing-matrix}
A^\dagger = A, \qquad (A\gamma ^\mu )^\dagger = A\gamma ^\mu \quad \mu =0, ...,3,
\ee
where $M^\dagger\equiv M^{*\,T}$ denotes the Hermitian conjugate of a matrix $M$. The existence of a nonzero matrix A satisfying (4) for a general set of Dirac matrices is proved in detail in Appendices \ref{uniqueness-proof} and \ref{uniqueness-affine}. Due to Eq. (\ref{hermitizing-matrix})$_1$, we define a Hermitian product between 4-vectors $u$ and $v$ by setting
\be\label{A-product}
(u ,v ) \equiv A_{\rho \nu } u ^{\rho*} v^\nu = u^\dagger A v.
\ee
The two properties (\ref{hermitizing-matrix}) are equivalent to the two following ones:
\be\label{hermitizing-A}
A_{\mu  \nu } = A_{ \nu \mu }^*, \qquad A_{\rho \nu } \left(\gamma^{\mu\, *} \right)^{\rho}  _{\ \sigma} = A_{\sigma  \rho }\left(\gamma^\mu \right)^{\rho  } _{\ \nu}  \qquad (\mu ,\nu  ,\sigma  \in \{0,...,3\}).
\ee 
The second property in (\ref{hermitizing-A}), in turn, means exactly \cite{A39} that each of the $\gamma ^\mu $ matrices is a Hermitian operator for the product (\ref{A-product}), that is,
\be\label{gamma-hermitian-A-product}
(\gamma ^\mu u,v) =(u,\gamma ^\mu v), \quad \mu =0, ...,3.
\ee
Let $\psi $ be a field defined on the spacetime manifold V, taking values in the vector space ${\sf C}^4$, and let us define the 4-current $j^\mu$ by \cite{A39}
\be\label{j-mu-standard}
j^\mu \equiv (\gamma ^\mu \psi ,\psi ) = A_{\rho \nu } \left(\gamma^{\mu\, *} \right)^{\rho}  _{\ \sigma}\psi ^{\sigma *} \psi ^{\nu }.
\ee
This may be rewritten as [cf. Eqs. (\ref{hermitizing-matrix}) and (\ref{A-product})]:
\be \label{j-mu-standard-matrix}
j^\mu =  \psi ^\dagger \gamma ^{\mu\,\dagger}\,A \,\psi = \psi^\dagger \,A\, \gamma ^{\mu}\,\psi .
\ee
Note that, until now, $g_{\mu \nu}$, $\gamma ^\mu$, and $A$, may depend on the spacetime point $X \in \mathrm{V}$. But, henceforth and until Sect. \ref{general-change}, we shall assume that spacetime is flat. Thus, there are Cartesian coordinates on V, such that the metric is the Poincar\'e-Minkowski metric, with component matrix
\be \label{g_munu-flat}
(\eta ^{\mu \nu})\equiv (\eta _{\mu \nu})^{-1}=(\eta _{\mu \nu})\equiv  \mathrm{diag}(1,-1,-1,-1).
\ee
We shall use a coordinate system $x^\mu$ derived from a Cartesian system by a linear transformation (an {\it affine system}), so that the flat metric has a general form $g^{\mu \nu}$, but is {\it constant}. In that case, also the $\gamma ^\mu$'s and $A$ are constant. We get then from the definition (\ref{j-mu-standard}), by using (\ref{gamma-hermitian-A-product}):
\be\label{drond_mu-jmu}
\partial _\mu j^\mu = (\gamma ^\mu \partial _\mu \psi ,\psi )+(\gamma ^\mu \psi ,\partial _\mu \psi )=(\gamma ^\mu \partial _\mu \psi ,\psi )+(\psi ,\gamma ^\mu \partial _\mu \psi ).
\ee
Let us assume now that the field $\psi $ obeys the Dirac equation
\footnote{\
In Sects. \ref{current} to \ref{uniqueness}, we shall consider the Dirac equation and its associated Hamiltonian operator in a fixed affine coordinate system in a flat spacetime.
} 
in the presence of an electromagnetic field characterized by the (real) potential $A_\mu$:
\be \label{Dirac-em}
i\gamma ^\mu (\partial_\mu +iqA_\mu)\psi  -m\psi =0 \qquad (\hbar =c=1).
\ee
Entering that into (\ref{drond_mu-jmu}) yields, using the sesquilinearity and (\ref{gamma-hermitian-A-product}):
\bea
\partial _\mu j^\mu & = & -(im\psi +iq\gamma ^\mu A_\mu\psi ,\psi )-(\psi ,im\psi +iq\gamma ^\mu A_\mu\psi )\nonumber\\
&=&im[(\psi ,\psi )-(\psi ,\psi )]+iq[(\gamma ^\mu A_\mu\psi ,\psi )-(\psi ,\gamma ^\mu A_\mu\psi )]\nonumber\\
&=& iq[(\gamma ^\mu A_\mu\psi ,\psi )-(\gamma ^\mu\psi , A_\mu\psi )].
\eea
That is, {\it the current is conserved:}
\be \label{conserved-current}
\partial _\mu j^\mu =0
\ee
in the presence of an electromagnetic field for a fully general choice of the Dirac matrices in a flat spacetime, expressed with affine coordinates.\\

We now show that, in a fixed affine coordinate system, {\it the current (\ref{j-mu-standard}) does not depend on the choice of the Dirac matrices}. Let $(\tilde {\gamma }^\mu)$ be {\it any other possible set,} thus satisfying the {\it same} anticommutation relation as does $(\gamma ^\mu)$:
\be \label{Clifford-tilde}
\tilde {\gamma} ^\mu \tilde {\gamma} ^\nu + \tilde {\gamma} ^\nu \tilde {\gamma}^\mu = 2g^{\mu \nu}\,{\bf 1}_4.
\ee
(Of course, metric $g^{\mu \nu}$ is unchanged, since we are not changing the coordinate system.) As shown by Pauli \cite{Pauli1933}, there exists a non-degenerate matrix $S$ such that the second set is obtained from the first one by the similarity transformation (which is a linear change of representation for the field $\psi $): 
\footnote{\
From Pauli's Fundamental Theorem (see Theorem 3 in Appendix \ref{uniqueness-proof}) one easily shows the existence of $S$  for a general metric, using Eqs. (\ref{psi-vector}) and (\ref{gamma-(^2_1)tensor}). 
}
\be \label{similarity-gamma}
\exists S \in {\sf GL}(4,{\sf C}):\qquad \tilde{\gamma} ^\mu =  S\gamma ^\mu S^{-1}, \quad \mu =0,...,3.
\ee
Moreover, $\psi $ obeys the Dirac equation (\ref{Dirac-em}) iff the similarity-transformed wave function,
\be \label{similarity-psi}
\tilde{\psi }\equiv  S\psi,
\ee
obeys the corresponding Dirac equation, involving matrices $\tilde{\gamma} ^\mu $. Using (\ref{similarity-gamma}) in (\ref{hermitizing-A}), one finds that the following matrix is hermitizing for the set $(\tilde{\gamma} ^\mu)$, i.e., after the similarity transformation:
\be \label{similarity-A}
\tilde {A}= (S^{-1})^\dagger A S^{-1}=(S^\dagger)^{-1} A S^{-1},
\ee
as is easily checked directly from (\ref{similarity-gamma}) and (\ref{hermitizing-matrix}). 
Hence, using (\ref{similarity-gamma})--(\ref{similarity-A}) in the definition (\ref{j-mu-standard-matrix}) of the current, we find:
\be \label{j-tilde}
\tilde {j}^\mu \equiv \tilde {\psi} ^\dagger \tilde {\gamma} ^{\mu\,\dagger}\,\tilde {A} \,\tilde {\psi}=(S\psi )^\dagger (S\gamma ^{\mu}S^{-1})^\dagger\,[(S^\dagger)^{-1} A S^{-1}] \,S\,\psi = j^\mu.\\
\ee
Thus, we have established the assertion that the current (\ref{j-mu-standard}) does not depend on the choice of the Dirac matrices.

\section{Hermitian Hamiltonian for a general set of Dirac matrices}\label{Hamiltonian}

Multiplying the Dirac equation (\ref{Dirac-em}) by $\gamma ^0$ on the left and using the anticommutation (\ref{Clifford}), one gets the Dirac equation in Schr\"odinger form:
\be \label{Schrodinger-general}
i \frac{\partial \psi }{\partial t}= \mathrm{H}\psi,\qquad (t\equiv x^0),
\ee
with 
\be \label{Hamilton-Dirac-em}
 \mathrm{H} \equiv  m\alpha  ^0 + \alpha ^j .(-i\partial _j) +q(A_0+\alpha ^j A_j),
\ee
and where 
\be \label{alpha}
\alpha ^0 \equiv \gamma ^0/g^{00}, \qquad \alpha ^j \equiv \gamma ^0\gamma ^j/g^{00}.
\ee
({\it We shall assume $g^{00}\ne 0$ throughout this paper.} Note that $p_j\equiv -i\partial _j$ commutes with $\alpha ^j$, because the latter is a constant matrix when using an affine coordinate system.) In order to study the hermiticity of the Hamiltonian (\ref{Hamilton-Dirac-em}), we shall use the existence of a matrix, say $B$, that is hermitizing for the $\alpha ^\mu $ 's:
\be\label{hermitizing-B}
B^\dagger = B, \qquad (B\alpha  ^\mu )^\dagger = B\alpha  ^\mu \quad \mu =0, ...,3.
\ee
Indeed, we prove in Appendix B the following result:\\

{\bf Theorem 6.}  {\it Consider the tensor Dirac theory, with transformation laws (\ref{psi-vector})-(\ref{gamma-(^2_1)tensor}). For any set of matrices $\gamma ^\mu$  satisfying the general anticommutation formula (\ref{Clifford}), there exists a hermitizing matrix $A$  for the matrices $\gamma ^\mu$.  The matrix  $A$ is nonsingular and unique, up to a real scale factor. Similarly, a nonsingular hermitizing matrix $B\equiv A\gamma ^0$ for the $\alpha ^\mu$'s exists and is unique, up to a real scale factor. If, furthermore, $g_{00}>0$ and the $3\times 3$ matrix $(g_{jk})\ (j,k=1,2,3)$ is negative definite, then $B\equiv A\gamma ^0$  is either a positive or negative definite matrix.  The sign of the matrix $A$ can be chosen such that $B\equiv A\gamma ^0$ is a positive definite matrix. }\\

We note that both conditions, i.e., that $g_{00}>0$ and that the matrix $(g_{jk})\ (j,k=1,2,3)$ be negative definite, must be true in any physically  admissible coordinate system \cite{L&L}. The coordinate systems for which these conditions are valid will be called {\it admissible}. According to the above Theorem, we can always introduce the Hermitian product
\be\label{B-product}
(u : v ) \equiv B_{\rho \nu } u ^{\rho*} v^\nu = u^\dagger B v.
\ee
As for Eq. (\ref{hermitizing-matrix}), it results from (\ref{hermitizing-B}) that each of the $\alpha  ^\mu $ matrices is a Hermitian operator {\it for the product (\ref{B-product}),} that is,
\be\label{alpha-hermitian-B-product}
(\alpha  ^\mu u: v) =(u: \alpha  ^\mu v), \quad \mu =0, ...,3.
\ee
If, moreover, the affine coordinate system considered is in fact an admissible one, then Theorem 6 shows that $B$ can be chosen to be a {\it positive} Hermitian matrix, so that the Hermitian product (\ref{B-product}) is positive in that case, $(u: u)>0$ if $u \ne 0$. Such a choice will be always assumed henceforth. Finally, it is well known that the operator $p_j \equiv -i\partial _j$ is Hermitian for the Hermitian product defined for scalar functions of space:
\be \label{Hermitian-general-scalar}
(a \mid  b ) \equiv \int_\mathrm{space} a({\bf x}) ^*\,b({\bf x}) \ \dd^ 3{\bf x} \qquad ({\bf x}\equiv (x^j)),
\ee
and it is easy to check that, when an operator $\mathcal{O}$ is extended from scalar functions to ones taking values in ${\sf C}^4$ by $\mathcal{O}.(\psi  ^\mu)\equiv (\mathcal{O}\psi  ^\mu)$, its adjoint for the product
\be \label{Hermitian-general-vector}
(\psi \parallel   \varphi  ) \equiv \int_\mathrm{space} (\psi ({\bf x}):\varphi({\bf x}))   \ \dd^ 3{\bf x}= B_{\rho \nu } (\psi  ^{\rho}\mid  \varphi ^\nu)
\ee
is the extension of the adjoint of $\mathcal{O}$ for the product (\ref{Hermitian-general-scalar})---so that $p_j$ is also Hermitian for the product (\ref{Hermitian-general-vector}). From this, and from (\ref{alpha-hermitian-B-product}), it follows that {\it the Dirac Hamiltonian (\ref{Hamilton-Dirac-em}) is a Hermitian operator for the Hermitian product (\ref{Hermitian-general-vector}),} which is a {\it positive} Hermitian product, if the coordinate system is an admissible one. \\

Also, the Hermitian product must give rise to a conserved probability density  $(\psi :\psi )$. Since $B=A\gamma ^0$, the conserved probability density in formula (\ref{j-mu-standard-matrix}) becomes $j^0=(\psi :\psi )$, thus  $j^0=(\psi :\psi )\geq 0$ in admissible coordinates in the Dirac theory with tensor transformation. Therefore, admissible coordinate systems play in that theory the role played in the DFW theory by the time-oriented tetrads. However, for the DFW theory, there are few studies on the hermiticity of the Hamiltonian in a generic coordinate system (even in general affine coordinates in a flat spacetime), except for Leclerc's work \cite{Leclerc2006}. Nevertheless, Leclerc assumes positive definiteness for the Hilbert space inner product, without proof. The positive definiteness of the Hilbert space inner product in general coordinates has not been addressed previously by other authors for the DFW theory, and certainly not for the Dirac theory with tensor transformation. \\

Let us investigate now, in a fixed admissible affine coordinate system, the influence of the choice of the set $(\gamma ^\mu)$ on the eigenvalues and eigenfunction expansions associated with the Dirac Hamiltonian (\ref{Hamilton-Dirac-em}). As in section \ref{current}, let $(\tilde {\gamma }^\mu)$ be another set of constant gamma matrices, satisfying the same anticommutation relation (\ref{Clifford}) as does $(\gamma ^\mu)$. Thus, the set $(\tilde {\gamma }^\mu)$ is obtained from the set $(\gamma ^\mu)$ by the similarity transformation (\ref{similarity-gamma}), from which it follows immediately that the matrices 
\be \label{alpha-tilde}
\tilde{\alpha} ^0 \equiv \tilde {\gamma} ^0/g^{00}, \qquad \tilde {\alpha} ^j \equiv \tilde {\gamma} ^0\tilde {\gamma} ^j/g^{00}
\ee
are obtained from the $\alpha ^\mu$'s (\ref{alpha}) by the same similarity transformation:
\be \label{similarity-alpha}
\tilde{\alpha } ^\mu =  S\alpha  ^\mu S^{-1}, \quad \mu =0,...,3.
\ee
Clearly, then, the Hamiltonian operator $\tilde{\mathrm{H}}$ corresponding to the set $(\tilde {\gamma }^\mu)$, which is defined by (\ref{Hamilton-Dirac-em}) with the matrices $\tilde{\alpha}^\mu$ in the place of the $\alpha ^\mu$'s, turns out to be simply
\be \label{similarity-H}
\tilde{\mathrm{H} }  =  S\mathrm{H} S^{-1}.
\ee
Moreover, matrix 
\be \label{similarity-B}
\tilde {B}= (S^{-1})^\dagger B S^{-1}=(S^\dagger)^{-1} B S^{-1},
\ee 
is a hermitizing matrix for the set $(\tilde{\alpha}^\mu)$. With the set $(\tilde {\gamma }^\mu)$, the relevant scalar product defined for wave functions $\tilde {\psi }$ and $\tilde {\varphi }$ is thus given by Eq. (\ref{Hermitian-general-vector}) with tildes. The respective wave functions exchange by $\tilde {\psi }= S\psi $, Eq. (\ref{similarity-psi}), since this is true for solutions of the respective Dirac equations. Using these two definitions and Eq.  (\ref{similarity-B}), it is straightforward to check, in the same way as for the invariance of the current [Eq. (\ref{j-tilde})], that the Hilbert space inner product (\ref{Hermitian-general-vector}) is invariant under similarity transformations:
\be \label{Hermitian-general-conserved}
(\tilde {\psi} \,\tilde {\parallel}\,   \tilde {\varphi}  )=(\psi \parallel   \varphi  ). 
\ee
Since Eq. (\ref{similarity-H}) implies that $\tilde{\mathrm{H} }  \,\tilde {\psi}=\widetilde{\mathrm{H}\psi}$, it follows then immediately that 
\be \label{Hamiltonian-matrix-conserved}
(\tilde{\mathrm{H} }  \,\tilde {\psi} \ \tilde {\parallel}\   \tilde {\varphi}  )=(\mathrm{H} \, \psi \parallel   \varphi  ). 
\ee
The first equation means that the mapping $\psi \mapsto S\psi $ is an isometry of the Hilbert space $\mathcal{H}$, relevant to the Dirac equation based on the set $(\gamma ^\mu)$, onto the Hilbert space $\tilde{\mathcal{H}}$, relevant to the Dirac equation based on the set $(\tilde {\gamma }^\mu)$. The second equation means that, if one chooses any Hilbert basis $(\psi _k)$ of the Hilbert space $\mathcal{H}$ [taken such that $\forall k, \ \psi _k \in \mathrm{Dom(H)}$], then the matrix of the Hamiltonian $\mathrm{H}$ in the basis $(\psi _k)$ is the same as the matrix of the Hamiltonian $\tilde {\mathrm{H}}$ in the Hilbert basis $(\tilde {\psi} _k)\equiv (S\psi _k)$ of the Hilbert space $\tilde{\mathcal{H}}$. In particular, the operators $\mathrm{H}$ and $\tilde {\mathrm{H}}$ have the same eigenvalues, the eigenfunction expansions of the states $\psi  \in \mathrm{Dom(H)}$ exchanging by the mapping $\psi \mapsto \tilde {\psi }\equiv S\psi $. In short, the quantum mechanics is fully unaffected by the choice of the set of Dirac matrices.\\

Note that the results of this section pertain to the tensor Dirac theory with the transformation behavior (\ref{psi-vector}) and (\ref{gamma-(^2_1)tensor}).  These results have not been proved for the DFW theory, except of course for Cartesian coordinates.  They do not apply to the genuine Dirac theory, except in the case of Cartesian coordinates.

\section{A uniqueness question}\label{uniqueness}

When discussing the current conservation in Sect. \ref{current} and the Hamiltonian in Sect. \ref{Hamiltonian}, we did not assume the uniqueness of the hermitizing matrix (\ref{hermitizing-matrix}), nor did we assume the uniqueness of the similarity transformation $S$ that transforms one set of Dirac matrices to another one, Eq. (\ref{similarity-gamma}). We do not need the uniqueness of the latter: any other possible transformation will lead to the same results, Eqs. (\ref{j-tilde}) and (\ref{Hermitian-general-conserved}), which express the absence of a dependence on the set of Dirac matrices. However, the non-uniqueness of the hermitizing matrix would mean that, for a {\it given} set of Dirac matrices, say $(\gamma ^\mu)$: i) there may exist several conserved currents $j$, each of them being given by Eq. (\ref{j-mu-standard}) with a different matrix $A$, that is hermitizing for the set $(\gamma  ^\mu)$; ii) there may exist several Hermitian products, each of them being given by Eqs. (\ref{Hermitian-general-vector}) and (\ref{B-product}) with a different matrix $B$, that is hermitizing for the set $(\alpha ^\mu)$. Note that this problem arises already for the standard set of Dirac matrices \cite{BjorkenDrell1964}, for which $A\equiv \gamma ^0$ turns out to be a hermitizing matrix, but is not necessarily the only one possible.\\

Obviously, the hermitizing matrix, say $A$, as characterized by property (\ref{hermitizing-matrix}), may be replaced by $\lambda A$ with any $\lambda \in {\sf R}^*$. From the explicit computation of Kofink \cite{Kofink1949}, it should follow that this is the only ambiguity which exists in the choice of the hermitizing matrix (denoted $B$ in the present work) for a set of alpha matrices $(\alpha ^\mu)$ satisfying the ``Euclidean" anticommutation (\ref{Clifford-h}) with $ h^{\mu \nu }=\delta ^{\mu \nu }$. To eliminate any doubt, this result is proved in Appendix \ref{uniqueness-proof}. Therefore, at least in the case of a {\it Cartesian} spacetime coordinate system, the Hermitian product (\ref{Hermitian-general-vector}) with (\ref{B-product}), with respect to which the Dirac Hamiltonian is a Hermitian operator, {\it is} unique up to a constant real factor---which is the best result that one can hope, and does not affect the spectrum of $\mathrm{H}$. It is also shown in Appendix \ref{uniqueness-proof} that the uniqueness of the hermitizing matrix is equally true for a set of gamma matrices $(\gamma ^\mu)$ in the case where the anticommutation relation involves the Poincar\'e-Minkowski metric $\eta ^{\mu \nu }$. That is, also the hermitizing matrix $A$ used in the definition (\ref{j-mu-standard}) of the current, is unique up to a non-zero real factor $\lambda $, at least in a Cartesian system. Thus, the current (\ref{j-mu-standard}) is also defined uniquely, up to a real factor---which is harmless.\\

In Appendix \ref{uniqueness-affine}, the existence of the hermitizing matrices $A$ and $B$, and their uniqueness up to a real scale factor, are extended to any affine coordinate system, using the tensor transformation (\ref{psi-vector})-(\ref{gamma-(^2_1)tensor}). For admissible affine coordinates, with an appropriate choice of sign, the positive definiteness of the hermitizing matrix  $B$ is established.  It follows that the scalar product  $(\psi \parallel \varphi )$ in (\ref{Hermitian-general-vector}), obtained from $(\psi : \varphi )$ in (\ref{B-product}), is a {\it positive} Hermitian product, with respect to which the Dirac Hamiltonian is a Hermitian operator. Furthermore, the scalar product $(\psi \parallel \varphi )$  is unique up to a positive real factor.  This assigns essentially a unique Hilbert space to each admissible affine coordinate system.  Acting in these Hilbert spaces, the Dirac Hamiltonian is a Hermitian operator in every admissible affine coordinate system

\section{Transforming the Dirac equation: the options} \label{transformation-options}

In this section, we shall recall some results of previous work \cite{A37}, adding a new observation. Let us investigate the transformation behaviour of the Dirac equation (\ref{Dirac-em}) under a linear coordinate change: 
\be \label{linearcoordinatechange}
x'^\mu =L^\mu_{\ \nu} x ^\nu, \qquad \mathrm{or}\quad x'=Lx \qquad (x\equiv (x^\mu )).
\ee
Let us restrict the consideration to those linear changes for which matrix $L$ belongs to some subgroup G of the group ${\sf GL}(4,{\sf R})$ of all possible linear transformations. One finds \cite{A37} that the covariance of the Dirac equation under a change (\ref{linearcoordinatechange}) depends on the existence of a {\it representation} (a group homomorphism) ${\sf S}$ of the group G into ${\sf GL}(4,{\sf C})$: for {\it any} pair $(\mathrm{G},{\sf S})$, the Dirac equation (\ref{Dirac-em}) is covariant, i.e., remains valid (with primes) in this same form (\ref{Dirac-em}) after any change (\ref{linearcoordinatechange}) with $L \in \mathrm{G}$, if we apply simultaneously the following changes to the wave function $\psi $ and to the matrices $\gamma ^\mu$:
\be \label{psi'}
\psi '(x')={\sf S}(L).\psi (x),
\ee
\be \label{gamma'}
\gamma '^\mu = L^\mu_{\ \nu} S\gamma ^\nu S^{-1}, \qquad S\equiv {\sf S}(L).
\ee
One may list three different possible choices for the pair $(\mathrm{G},{\sf S})$:
\begin{itemize}
\item {\bf i}. \ G $={\sf SO}^+(1,3)$, the proper orthochronous Lorentz group, with ${\sf S}$ being the spinor representation. This is the standard choice, in fact Dirac's original choice, briefly discussed at the beginning of the present paper.

\item {\bf ii}. \ G $={\sf GL}(4,{\sf R})$, with ${\sf S}$ being the identity representation defined by ${\sf S}(L)=L \ \forall L \in \mathrm{G}$. This is the ``$\psi =$ vector and $(\gamma ^\mu )=(^2_1)$ tensor" transformation behaviour \cite{A37}, Eqs. (\ref{psi-vector})--(\ref{gamma-(^2_1)tensor}) in the present work. It will be designated shortly as ``the tensor transformation" of the Dirac equation.

\item {\bf iii}. \ G $={\sf GL}(4,{\sf R})$, with ${\sf S}$ being the trivial representation defined by ${\sf S}(L)={\bf 1}_4 \ \forall L \in \mathrm{G}$. (The possibility of this choice had not been noted in the previous work \cite{A37}.) Thus Eqs. (\ref{psi'})--(\ref{gamma'}) become
\be \label{psi'-S=1}
\psi '(x')=\psi (x),
\ee
\be \label{gamma'-S=1}
\gamma '^\mu = L^\mu_{\ \nu} \gamma ^\nu .
\ee
This is nothing else than the transformation behaviour associated \cite{BrillWheeler1957+Corr} with the standard gravitational extension of the Dirac equation \cite{Weyl1929b,Fock1929b}, here named the Dirac-Fock-Weyl (DFW) equation.

\end{itemize}

\noindent The two last possibilities, in contrast with the first one, are defined for any linear coordinate change. This is the reason why, after introducing some covariant derivative, they extend to general coordinate changes. This is well known for the DFW equation, associated with choice {\bf iii}, and it will be shown in Sect. \ref{general-change} for choice {\bf ii}.

\section{Tensor transformation of the Dirac equation}\label{linear-change} 

As noted in the Introduction, it is a closed set of equations {\it together with their transformation behaviour} that makes a definite physical theory. It will be clear now that the Dirac equation with tensor transformation [let us call it ``the tensor Dirac theory," defined by the transformation scheme {\bf ii} in the foregoing section] is physically equivalent to the Dirac equation with spinor transformation [``the spinor Dirac theory," defined by scheme {\bf i}], in the domain of validity of the latter, i.e., in inertial frames (with Cartesian coordinates).
\footnote{\ 
We consider the e.m. potential $A_\mu $ as given, hence the Dirac equation (together with the relevant boundary conditions, of course) is a closed system.
} 
To show this, we may choose the inertial frame as we wish, since each of the two theories is covariant under Lorentz transformations (Sect. \ref{transformation-options}). But we were not allowed {\it a priori} to select the same set $(\gamma ^\mu )$ of Dirac matrices, since the tensor Dirac theory does not leave the $\gamma ^\mu $'s invariant. However, we now know that the choice of the set $(\gamma ^\mu )$ is fully immaterial, since the probability current $j$, as well as all scalar products $(\psi \parallel   \varphi  )$ and transition amplitudes $(\mathrm{H} \, \psi \parallel   \varphi  )$, are invariant under any similarity transformation; i.e., any change of the set $(\gamma ^\mu )$ (associated with a linear change of representation of the field $\psi $)  [Sects. \ref{current} and \ref{Hamiltonian}]. Therefore, in our selected inertial frame, we {\it may} after all take the same set $(\gamma ^\mu )$ for the spinor and the tensor Dirac theories. Thus, the two theories being associated with just the same equation (with the same matrix coefficients) in a given inertial frame, their equivalence is obvious.\\

As a complementary check of the consistency of the tensor Dirac theory, let us investigate the transformation properties of a few additional objects under coordinate changes, according to the tensor Dirac theory. Until Section \ref{general-change}, the  Hamiltonian (21) is restricted to a flat spacetime with certain affine coordinates. However, the tensor transformation of the wave function and the Dirac matrices extends to general coordinate changes, and this is also true for the transformation of the objects studied below.
\begin{itemize}
\item The {\it hermitizing matrix} $A=(A_{\rho \nu })$ may be characterized by Eq. (\ref{hermitizing-A}). Since $\gamma ^{\mu \rho }_\nu \equiv (\gamma ^{\mu})^ \rho _{\ \nu}$ is a $(^2 _1)$ tensor, it follows that (\ref{hermitizing-A}) is covariant if we transform $A_{\rho \nu} $ as a $(^0 _2)$ tensor. (This does not depend on the uniqueness of the hermitizing matrix, which is studied in Appendices \ref{uniqueness-proof} and \ref{uniqueness-affine}.) In other words, $A=(A_{\rho \nu })$ is a hermitizing matrix in the coordinates $x^\mu$ iff
\be \label{A-(0 2)tensor}
A'= \left(L^{-1}\right)^T A L^{-1}, \qquad L^\mu_{\ \nu} \equiv \frac{\partial x'^\mu }{\partial x^\nu }
\ee
is a hermitizing matrix in the coordinates $x'^\mu$. (It preserves the hermiticity: $A^{*\,T}=A$ and (\ref{A-(0 2)tensor}) imply that $A'^{*\,T}=A'$.)

\item The {\it current} $j$, with $j^\mu\equiv A_{\rho \nu } \gamma ^{\mu \rho *}_\sigma  \psi ^{\sigma *} \psi ^{\nu }$ [Eq. (\ref{j-mu-standard})], is therefore a vector, as it is also in the spinor Dirac theory.

\item The {\it charge conjugation matrix} may be defined to be a matrix $C=(C^\rho _{\ \nu })$ such that
\footnote{\
The matrix noted $C$ here is often denoted as $C\gamma ^0$.  E.g., Eq. (\ref{charge-conjugation-matrix}) is equivalent to Eq. (5.4) on p. 67 in Bjorken and Drell \cite{BjorkenDrell1964}, with here $C$ in the place of $C\gamma ^0$ in Ref. \cite{BjorkenDrell1964}.
}
\be\label{charge-conjugation-matrix}
 C\gamma ^{\mu \,*} = -\gamma ^\mu C, \quad \mu =0, ...,3,
\ee
hence in the tensor Dirac theory:
\be
C^\rho _{\ \nu }\gamma ^{\mu \nu  *}_\sigma =-\gamma ^{\mu \rho }_\nu C^\nu  _{\ \sigma  },
\ee
which shows that $(C^\rho _{\ \nu })$ is indeed a $(^1 _1)$ tensor, i.e.,
\be\label{C-tensor}
C'=L C L^{-1}.
\ee
In other words, $C$ commutes with all coordinate transformations.

\item The ``{\it gamma-five matrix,"} defined in a Poincar\'e-Minkowski spacetime by
\be\label{gamma5}
\gamma ^5\equiv i \gamma ^0 \gamma ^1 \gamma ^2 \gamma ^3,
\ee
may be equivalently defined, in a general coordinate system in a general spacetime with metric $\Mat{g}\equiv (g_{\mu \nu })$, by
\be\label{gamma5-general}
\gamma ^5\equiv \frac{i}{24}\ e_{\mu \nu \rho \sigma }\gamma ^\mu  \gamma ^\nu  \gamma ^\rho  \gamma ^\sigma ,
\ee
with the following $(^0 _4)$ tensor (for transformations with $\mathrm{det}(L)>0$):
\be\label{e-mu-nu-rho-sigma}
e_{\mu \nu \rho \sigma } \equiv \sqrt{\abs{g}}\ \epsilon _{\mu \nu \rho \sigma },
\ee
where $g\equiv \mathrm{det}(g_{\mu \nu })$ and $\epsilon _{\mu \nu \rho \sigma }$ is the signature of the permutation $(\mu \nu \rho \sigma )$ of $\{0,...,3\}$. We have
\be \label{D-tensor}
(\gamma ^\mu  \gamma ^\nu  \gamma ^\rho  \gamma ^\sigma)^\tau   _{\ \phi  } =\gamma ^{\mu \tau  }_\chi \gamma ^{\nu \chi  }_\omega \gamma ^{\rho \omega }_\zeta \gamma ^{\sigma  \zeta  }_\phi \equiv D^{\mu \nu \rho \sigma \tau }_\phi,
\ee
which is a $(^5 _1)$ tensor. It follows thus from (\ref{gamma5-general}) and (\ref{D-tensor}) that  $T ^{\tau}   _{\ \phi  } \equiv (\gamma ^5)^\tau   _{\ \phi  } $ is a $(^1 _1)$ tensor, or
\be\label{gamma5-tensor}
\gamma'^5=L\gamma ^5 L^{-1}.
\ee
Thus, $\gamma ^5$ commutes with all coordinate transformations, as does $C$.

\item The previous results allow us to study the transformation of the {\it Hestenes tensor fields} \cite{Hestenes1967}. In the form given by Takahashi \cite{Takahashi1983} and extended to a general spacetime by Reifler and Morris \cite{ReiflerMorris1999,ReiflerMorris2000,ReiflerMorris2005}, these are a scalar field $s$ and a tetrad field $e_K^\mu $ given by 
\footnote{\
In fact, this definition extends that of Reifler and Morris \cite{ReiflerMorris2005}, in that it uses the the general hermitizing matrix $A$, thus in Ref. \cite{ReiflerMorris2005} we would define $\bar {\psi }\equiv \psi ^\dagger A $ instead of $\bar {\psi }\equiv \psi ^\dagger \gamma ^0 $. Matrix $\gamma ^0$ is a hermitizing matrix for the standard set \cite{BjorkenDrell1964} of Dirac matrices.
}
\bea\label{s-eK}
s \equiv  & (\psi ,\psi )- (\psi ,\gamma ^5 \psi),&\\
e_K \equiv  &\abs{s}^{-1}J_K ,&
\eea
where the currents $J_K$ $(K=0,...,3)$ are defined by 
\be\label{JK}
J_K^\mu \equiv \mathrm{Re} \left(\psi ,i\gamma ^\mu \tau _K \psi\right)  ,
\ee
with
\be\label{tauK-1}
\tau _0\psi \equiv  -i\psi, \qquad \tau _1\psi   \equiv iC\psi ^*,
\ee
\be\label{tauK-2}
\tau _2\psi   \equiv  C\psi ^*, \qquad \tau _3\psi   \equiv i\gamma ^5 \psi.
\ee

It results from (\ref{psi-vector}), (\ref{A-(0 2)tensor}) and (\ref{gamma5-tensor}) that $s$ is indeed an invariant scalar, and it results from (\ref{psi-vector}), (\ref{C-tensor}) and (\ref{gamma5-tensor}) that each $\tau _K \psi$ $(K=0,...,3)$ is a spacetime vector. (Thus, $\psi $ itself being also a vector, each $\tau _K$ commutes with all coordinate transformations, as also do $C$ and $\gamma ^5$.) It then follows that each of the currents $J_K$ $(K=0,...,3)$, and so also each of the $e_K$'s, is a spacetime vector.

\end{itemize}
We end this section by noting that, in the past, ``the tensor Dirac theory," or ``the tensor formulation of Dirac theory," has designated formulations of the Dirac theory in terms of the Hestenes tensor fields $s$ and $e_K^\mu $. Those may be defined independently of the transformation of the wave function and the gamma matrices, and it was actually the spinor transformation scheme that was used previously when discussing the Hestenes tensor fields \cite{Hestenes1967, ReiflerMorris1999, ReiflerMorris2000, ReiflerMorris2005}. Indeed all bispinor observables, such as the electric and chiral currents, energy-momentum and spin-polarization tensors, as well as the bispinor Lagrangian, can be expressed in terms of Hestenes' scalar and tetrad fields \cite{ReiflerMorris2003}.

\section{Conclusion: tensor Dirac theory in general coordinates}\label{general-change}
While concluding their well-known paper about the Dirac-Fock-Weyl theory, Brill and Wheeler \cite{BrillWheeler1957+Corr} asked: ``What is there about the geometry of space which is not already adequately covered by ordinary scalars, vectors, and tensors of standard tensor analysis?" and they noted that ``spinors allow one to describe rotations at one point in space completely independently of rotations at all other points in space. Fully to see at work this machinery of independent rotations at each point in space, we do best to consider the spinor field in a general curved space [...]. But the deeper part of such rotations in the description of nature is still mysterious."\\

In the present paper, after having established the matrix-representation independence of the Dirac theory, we have been able to show that the {\it tensor Dirac theory} is quantum-mechanically fully equivalent to the genuine Dirac theory, involving spinor transformation of the wave function and Lorentz-invariant gamma matrices. Here, we mean by ``tensor Dirac theory," the {\it usual} Dirac equation with {\it vector wave function} and with the set of the components of the gamma matrices building a {\it third-order tensor,} Eqs. (\ref{psi-vector}) and (\ref{gamma-(^2_1)tensor}) \cite{A37}. This physical equivalence is not an obvious result, as the Dirac theory has been steadily associated, since its discovery, to the spinor transformation. Thus, the answer to Brill and Wheeler's question seems to be that in fact, at least in a flat spacetime, there is nothing to add to ``standard tensor analysis."  \\

In addition to its greater simplicity, the tensor Dirac equation has the advantage that, at least in a flat spacetime, it extends immediately to a general coordinate system: since all objects are tensors, we merely have to replace the partial derivatives $\partial _\mu $ in the Dirac equation (\ref{Dirac-em}) by the covariant derivatives $D_\mu $ with respect to the Levi-Civita connection, thus defining 
\be\label{D_mu-psi}
(D_\mu \psi )^\nu =\partial _\mu\psi ^\nu +\Gamma ^\nu _{\rho \mu }\psi ^\rho,
\ee
the $\Gamma ^\nu _{\rho \mu }$'s being the Christoffel symbols, associated with metric $\Mat{g}$. In the case that $\Mat{g}$ is {\it flat}, indeed, we know that (\ref{D_mu-psi}) is the {\it only} covariant derivative that behaves as a {\it tensor} and coincides with $\partial _\mu \psi^\nu $ in Cartesian coordinates \cite{Doubrovine1982}. 
\footnote{\ 
For the DFW theory and its generalizations to more general spacetimes, $\psi ^\nu $   is not a vector, and  $(D_\mu \psi )^\nu$ is defined to be the ``spinor derivative," which does not behave as a tensor with respect to both indices $\mu $  and $\nu $ \cite{BrillWheeler1957+Corr,Poplawski2007}.  
}
This opens a new, more direct possibility to study problems in non-inertial (e.g. rotating) frames for fermions, as compared to using the Dirac-Fock-Weyl equation. On the other hand, in the case that the space-time metric $\Mat{g}$ is {\it curved}, we {\it may} keep the definition (\ref{D_mu-psi}) and we thus obtain a gravitational extension of the tensor Dirac equation, which obeys the equivalence principle \cite{A39}. But another extension is possible, that leads to a theory with a physically-preferred reference frame \cite{A39}. Both of these gravitational extensions of the tensor Dirac theory have still a tentative status, because two important points remain to be studied: the possibility of defining i) a conserved current and ii) a Hermitian scalar product making the Hamiltonian operator Hermitian. (These two questions are being investigated in detail in a forthcoming work \cite{A42}.) However, we believe that they become more interesting, now that it has been proved that the tensor Dirac equation itself is equivalent to the genuine Dirac equation.

\appendix
\section{Uniqueness proof of hermitizing matrices with the Poincar\'e-Minkowski metric}\label{uniqueness-proof}

\noindent {\bf Theorem 1.}  {\it Any matrix that commutes with all four gamma matrices  $\gamma ^{\sharp \mu} $ in the Dirac representation is a complex scalar multiple of the identity matrix ${\bf 1}_4$.  Also, any matrix that commutes with all four alpha matrices $\alpha ^{\sharp \mu}$  in the Dirac representation is a complex scalar multiple of the identity matrix. } \\

{\it Proof.}  Any matrix, which commutes with every gamma (alpha) matrix, commutes with the entire Dirac algebra generated by products of gamma (alpha) matrices.  As is well-known (see e.g. ref. \cite{BjorkenDrell1964}), the algebra $\mathcal{D}$ generated by the gamma matrices contains every $4\times 4$ complex matrix. This is hence also true for the algebra $\mathcal{D}'$ generated by the alpha matrices, since $\gamma ^{\sharp 0}=\alpha ^{\sharp 0}$ and $\gamma ^{\sharp j}=\alpha ^{\sharp 0}\alpha ^{\sharp j}$ belong to $\mathcal{D}'$. But a $4\times 4$ matrix which commutes with all complex $4\times 4$ matrices is a multiple of the identity.  Q.E.D.\\

\noindent {\bf Theorem 2.}  {\it Any hermitizing matrix $B^\sharp $ for the alpha matrices  $\alpha ^{\sharp \mu}$ in the Dirac representation is a real scalar multiple of the identity matrix.}\\

{\it Proof.}  Since each  $\alpha ^{\sharp \mu}$ in the Dirac representation is Hermitian, and  $B^\sharp $ is hermitizing for  $\alpha ^{\sharp \mu}$,
\be 
\left( B^\sharp \alpha ^{\sharp \mu}\right) ^\dagger = \alpha ^{\sharp \mu} B^\sharp = B^\sharp \alpha ^{\sharp \mu},
\ee
so that  $B^\sharp $ commutes with all four matrices  $\alpha ^{\sharp \mu}$.  Therefore   $B^\sharp $ is a scalar multiple of the identity matrix.  Since  $B^\sharp $ is Hermitian, the scalar must be real.  Q.E.D.\\
    
\noindent {\bf Theorem 3}  {\it (Pauli's Fundamental Theorem).  For any set of matrices  $\gamma ^\mu$ satisfying the anticommutation formula with the Poincar\'e-Minkowski metric:
\be \label{Clifford-flat}
\gamma ^\mu \gamma ^\nu + \gamma ^\nu \gamma ^\mu = 2\eta ^{\mu \nu}\,{\bf 1}_4, \quad \mu ,\nu \in \{0,...,3\},
\ee
there is a similarity transformation (\ref{similarity-gamma}), i.e., a change of representation, that takes $\gamma ^\mu$  to $\gamma  ^{\sharp \mu}$ , and takes  $\alpha  ^\mu$  to  $\alpha ^{\sharp \mu}$, giving the Dirac representation of the gamma and alpha  matrices.  The similarity transformation is unique up to a nonzero complex scalar multiple of the identity matrix.}\\

{\it Proof.} The existence is a well-known result, obtained by Pauli \cite{Pauli1933}. Note that, if a similarity transformation (\ref{similarity-gamma}) takes  $\gamma ^\mu$  to  $\gamma  ^{\sharp \mu}$, it follows from the general definition (\ref{alpha}) of the alpha matrices that it also transforms  $\alpha  ^\mu$  into  $\alpha ^{\sharp \mu}$. Let us prove the uniqueness. Let  $S$ and $T$  be similarity transformations both taking $\gamma ^\mu$  to  $\gamma  ^{\sharp \mu}$.  Then $TS^{-1}$ commutes with each  $\gamma  ^{\sharp \mu}$, and therefore, using Theorem 1, is a nonzero complex scalar multiple of the identity matrix.  Q.E.D. \\

\noindent {\bf Theorem 4.}  {\it  For any set of matrices  $\gamma ^\mu$ satisfying the anticommutation formula (\ref{Clifford}) with a general metric, let $A$  be any matrix that is hermitizing for the matrices  $\gamma ^\mu$.  Then 
\be
B\equiv A\gamma ^0
\ee 
is hermitizing for the alpha matrices  $\alpha  ^\mu$ (\ref{alpha}). Conversely, if $B$ is any matrix that is hermitizing for the matrices  $\alpha  ^\mu$, then $A=B\alpha ^0$  is hermitizing for the gamma matrices  $\gamma ^\mu$.      }\\  

{\it Proof.} Since  $A$ is hermitizing for the $\gamma ^\mu$'s, we have
\be
B^\dagger = \left(A\gamma ^0\right)^\dagger = A\gamma ^0 =B,
\ee
so that  $B$ is a Hermitian matrix. Let us show that $B$ is a hermitizing matrix for the $\alpha  ^\mu$'s. Using also the definition (\ref{alpha}) and the anticommutation (\ref{Clifford}), we get
\be
\left(B \alpha ^0\right)^\dagger = \left(A\gamma ^0 \gamma ^0/g^{00}\right)^\dagger =A^\dagger= A = A \gamma ^0 \gamma ^0/g^{00} = B\alpha ^0,
\ee
\be
\left(B \alpha ^j\right)^\dagger = \left(A\gamma ^0 \gamma ^0\gamma ^j/g^{00}\right)^\dagger =\left(A\gamma ^j\right)^\dagger= A\gamma ^j = A \gamma ^0 \gamma ^0\gamma ^j/g^{00} = B\alpha ^j,
\ee 
which shows that $B=A\gamma ^0$  is hermitizing for the alpha matrices $\alpha  ^\mu$.  The converse assertion is proved similarly.  Q.E.D.\\       

\noindent {\bf Theorem 5.}  {\it For any set of matrices  $\gamma ^\mu$ satisfying the anticommutation formula with metric $\eta ^{\mu \nu }$, Eq. (\ref{Clifford-flat}), there exists a hermitizing matrix  $A$ for the matrices $\gamma ^\mu$.  The matrix  $A$ is nonsingular and unique, up to a real scale factor.  Moreover,  matrix $A\gamma ^0$ is hermitizing for the alpha matrices $\alpha  ^\mu$.  The matrix  $A\gamma ^0$ is either positive definite or negative definite.  The sign of the matrix  $A$  can be chosen so that $A\gamma ^0$  is a positive definite matrix.\\  

Furthermore, let  $B$ be any matrix that is hermitizing for the matrices $\alpha  ^\mu$.  Then, $B$  is a real scalar multiple of  $A\gamma ^0$.}\\

{\it Proof.}  Using Theorem 3, there is a similarity transformation  $S$ that takes $\gamma ^\mu$  to  $\gamma ^{\sharp \mu}$, and takes $\alpha   ^\mu$  to  $\alpha  ^{\sharp \mu}$, giving the Dirac representation of the gamma and alpha  matrices.  Then   $A=S^\dagger \gamma ^{\sharp 0}S$ is a nonsingular hermitizing matrix for the matrices $\gamma ^\mu$ [cf. Eq. (\ref{similarity-A})].  Now let $A'$ be another nonzero hermitizing matrix for the $\gamma ^\mu$'s.  Let the similarity transformation $S$  take  $A'$  to $A^\sharp $, in the sense of Eq. (\ref{similarity-A}).  Thus,  $A^\sharp $ is hermitizing for the $\gamma ^{\sharp \mu}$'s.  Then, by Theorem 4, $A^\sharp \gamma ^{\sharp 0}$  is hermitizing for the $\alpha  ^{\sharp \mu}$'s.  Using Theorem 2, we get  $A^\sharp \gamma ^{\sharp 0}=\lambda {\bf 1}_4$, where $\lambda $  is a nonzero real scalar.  Hence,  $A^\sharp =\lambda \gamma ^{\sharp 0}$.  Therefore,
\be\label{101}
A'=S^\dagger A^\sharp S =\lambda S^\dagger \gamma ^{\sharp 0} S =\lambda A,
\ee
which shows that the matrix  $A'$ is a  nonzero real scalar multiple of $A$.  Thus, the matrix $A$  is nonsingular and unique, up to a real scale factor. \\  

Now let $B$  be any nonzero matrix that is hermitizing for the alpha matrices $\alpha   ^\mu$.  Let the above-mentioned similarity transformation  $S$ take  $B$  to $B^\sharp$, in the sense of Eq. (\ref{similarity-B}).  Then $B^\sharp$  is hermitizing for the $\alpha  ^{\sharp \mu}$'s.  Using Theorem 2, we get $ B^\sharp=\mu {\bf 1}_4$ where again $\mu $  is a nonzero real scalar.  Therefore,
\be\label{102}
B=S^\dagger B^\sharp S =\mu  S^\dagger S,
\ee
which, from the uniqueness of $S$ as stated in Theorem 3, establishes that the matrix  $B$ is unique up to a  nonzero real scalar factor.  Furthermore, since $\mu $  is a nonzero real scalar, formula (\ref{102}) shows that $B$  is either a positive definite or a negative definite matrix. \\

Using the equation  $\gamma ^{\sharp 0}=S\gamma ^0 S^{-1}$ in formula (\ref{101}), we see from formula (\ref{102}) that $B=\mu A\gamma ^0$ .  Thus, since  $B$ is a nonzero matrix,  $A\gamma ^0$ is a nonzero real scalar multiple of $B$, and hence hermitizing for the alpha matrices  $\alpha ^\mu$, and is either a positive definite or a negative definite matrix.  Clearly, the sign of the matrix  $A$  can be chosen so that  $A\gamma ^0$ is a positive definite matrix. The proof is complete.

\section{Uniqueness of hermitizing matrices with a general metric in the tensor Dirac theory}\label{uniqueness-affine}
{\bf Theorem 6.}  {\it Consider the tensor Dirac theory, with transformation laws (\ref{psi-vector})-(\ref{gamma-(^2_1)tensor}). For any set of matrices $\gamma ^\mu$  satisfying the general anticommutation formula (\ref{Clifford}), there exists a hermitizing matrix $A$  for the matrices $\gamma ^\mu$.  The matrix  $A$ is nonsingular and unique, up to a real scale factor. Similarly, a nonsingular hermitizing matrix $B\equiv A\gamma ^0$ for the $\alpha ^\mu$'s exists and is unique, up to a real scale factor. If, furthermore, $g_{00}>0$ and the $3\times 3$ matrix $(g_{jk})\ (j,k=1,2,3)$ is negative definite, then $B\equiv A\gamma ^0$  is either a positive or negative definite matrix.  The sign of the matrix $A$ can be chosen such that $B\equiv A\gamma ^0$ is a positive definite matrix. }\\

\indent The proof of Theorem 6 uses the following\\

{\bf Lemma.} {\it If $g_{00}>0$ and the $3\times 3$ matrix $(g_{jk})\ (j,k=1,2,3)$ is negative definite, then there is a linear coordinate transformation $L=(L^\mu _{\ \,\nu })$ of the form 
\be\label{L^0_k=0}
\left (\begin{array}{cc} L^0_{\ \,0} \quad L^0_{\ \,k} \\ L^j_{\ \,0}\quad L^j_{\ \,k} \end{array} \right)=\left (\begin{array}{cc} \lambda \quad 0 \\ \lambda^j \quad L^j_{\ \,k} \end{array} \right),
\ee
that takes $(\eta _{\mu \nu })$  to $(g _{\mu \nu })$, i.e., 
\be\label{eta-to-g}
g _{\mu \nu }=\eta _{\alpha \beta  }\, L^\alpha _{\ \,\mu } \,L^\beta  _{\ \,\nu },
\ee
with $\lambda >0$. The  $4\times 4$ matrix $L$ can be chosen as a unique extension of a $3\times 3$ matrix $l \equiv (L^j_{\ \,k})$, that takes  $(\eta _{jk})$  to $(g_{j k })$.  Furthermore, $M\equiv L^{-1}$  takes $(\eta ^{\mu \nu })$ to $(g^{\mu \nu })$, i.e., $g ^{\mu \nu }=\eta ^{\alpha \beta  }\, M^\mu  _{\ \,\alpha  } \,M^\nu   _{\ \,\beta  }$, and has the form 
\be\label{M^0_k=0}
\left (\begin{array}{cc} M^0_{\ \,0} \quad M^0_{\ \,k} \\ M^j_{\ \,0}\quad M^j_{\ \,k} \end{array} \right)= \left (\begin{array}{cc} \lambda^{-1} \qquad \qquad \qquad \qquad 0 \\ - \sum_{\substack{k=1}}^{\substack{3}} \lambda^{-1}\left(l^{-1}\right)^j_{\ \,k}\lambda^k \quad \left(l^{-1}\right)^j_{\ \,k} \end{array} \right).
\ee
Thus, in particular, we have $M^0_{\ \,0}>0$, and $M^0_{\ \,k}=0$.}\\

{\it Proof.}  We have to find $L$ of the form (\ref{L^0_k=0}), such that  Eq. (\ref{eta-to-g}) is satisfied.  Since the $3\times 3$ matrices $(\eta _{jk})$  and $(g_{jk})$ are both negative definite, there is a linear transformation $l \equiv (L^j_{\ \,k})$ of the spatial coordinates, that takes $(\eta _{jk})$ to $(g_{jk})$:
\be\label{g_jk}
g _{jk}=\eta _{mn }\, L^m _{\ \,j } \,L^n  _{\ \,k }=-L^m _{\ \,j } \,L^m  _{\ \,k }.
\ee
Thus, the components $(\mu =j \in \{1,2,3\}, \nu =k\in \{1,2,3\})$ of Eq. (\ref{eta-to-g}) are satisfied. Since the $3\times 3$ matrix $l$ is nonsingular, we can then uniquely solve for $\lambda ^j\equiv L^j_{\ \,0}$ in the components $(\mu =j \in \{1,2,3\}, \nu =0)$ of Eq. (\ref{eta-to-g}), by imposing the condition $L^0_{\ \,j}=0 \ (j=1,2,3)$:
\be
g_{j0}=-\sum_{\substack{k=1}}^{\substack{3}} L^k_{\ \,j}\lambda^k.
\ee
The components $(\mu =0, \nu =j \in \{1,2,3\})$ of Eq. (\ref{eta-to-g}) are then automatically satisfied. Now, since $g_{00}>0$, we can uniquely solve for $\lambda>0 $  from the last component $(\mu = \nu =0)$ of Eq. (\ref{eta-to-g}):
\be
g_{00}=\lambda ^2-\sum_{k=1}^{3} \left(\lambda^k\right)^2.
\ee
Owing to (\ref{eta-to-g}), $L$ is the matrix of the transformation from the starting coordinates $x^\mu$ to the coordinates $x'^\alpha $, in which the metric is $g'_{\alpha \beta  }=\eta _{\alpha \beta  }$. By construction, $L$ satisfies Eq. (\ref{L^0_k=0}), thus
\be
x'^0=L^0_{\ \,0}x^0, \qquad x'^j=L^j_{\ \,0}x^0 + \sum_{k=1}^{3} L^j_{\ \,k}x^k,
\ee
and by inversion:
\be
x^0=\left(L^0_{\ \,0}\right)^{-1}x'^0, \qquad x^j= \sum_{k=1}^{3} \left(l^{-1}\right)^j_{\ \,k} [x'^k-L^k_{\ \,0}\left(L^0_{\ \,0}\right)^{-1}x'^0],
\ee
whence the form (\ref{M^0_k=0}) for matrix $M$. Q. E. D.\\

{\it Proof of Theorem 6.}  Choose a linear coordinate transformation, with matrix $L$,  that takes $g^{\mu \nu } $ to $\eta ^{\mu \nu } $.  By Eq. (\ref{gamma-(^2_1)tensor}), $L$ takes $\gamma ^\mu$ to $\gamma'^\mu$, and the anticommutation relation transforms covariantly \cite{A37}, thus    
\be\label{103}
\gamma'^\mu \gamma'^\nu + \gamma'^\nu \gamma'^\mu = 2\eta ^{\mu \nu}\,{\bf 1}_4, \quad \mu ,\nu \in \{0,...,3\}.
\ee
By Theorem 5, there exists a nonsingular hermitizing matrix $A'$ for the matrices $\gamma'^\mu$.  As stated with Eq. (\ref{A-(0 2)tensor}), $A\equiv  L^T A' L$ is a nonsingular hermitizing matrix for the matrices $\gamma^\mu$.  Moreover, if $\tilde{A}$ is another nonzero hermitizing matrix for the gamma matrices $\gamma^\mu$, then $L$ takes $\tilde{A}$ to $\tilde{A}'\equiv \left(L^{-1}\right)^T \tilde{A} L^{-1}$, which is another nonzero hermitizing matrix for the matrices $\gamma'^\mu$.  Since $\tilde{A}'$ is a nonzero real scalar multiple of $A'$ by Theorem 5, $\tilde{A}=  L^T \tilde{A}' L$ is a nonzero real scalar multiple of $A$.  Thus, the matrix  $A$ is nonsingular and unique, up to a real scale factor.\\       

As we have just shown, a nonsingular hermitizing matrix $A$ does exist for the gamma matrices $\gamma ^\mu$. By Theorem 4,  $A\gamma ^0$ is a hermitizing matrix for the alpha matrices $\alpha ^\mu$.  Since $\alpha ^0=(\gamma ^0)^{-1}$ from formula (\ref{alpha}), it follows that the matrix  $\gamma ^0$ is nonsingular. Thus, there exists a nonsingular hermitizing matrix for the alpha's, namely $ A\gamma ^0$. Theorem 4 also states that, if $B$ is any matrix that is hermitizing for the matrices  $\alpha  ^\mu$, then $A=B\alpha ^0$  is hermitizing for the gamma matrices  $\gamma ^\mu$. The uniqueness (up to a real scale factor) of $B$ follows thus from that of $A$, proved in the above paragraph.\\

If now $g_{00}>0$ and the $3\times 3$ matrix $(g_{jk})\ (j,k=1,2,3)$ is negative definite, $L$ as prescribed above and its inverse matrix $M=L^{-1}$ can be chosen as in the Lemma. In particular, we have $M^0_{\ \,k}=0$. From this, using also (\ref{gamma-(^2_1)tensor}) and the foregoing relation $A\equiv  L^T A' L$, we obtain the transformation for the matrix $A\gamma ^0$ as follows:
\be\label{105}                         
A\gamma ^0 = L^T A'L M^0_{\ \,\nu } M \gamma'^\nu M^{-1}= M^0_{\ \,0} L^T A'\gamma'^0 L,
\ee
where $M^0_{\ \,0}>0$ by the Lemma. With an appropriate choice of sign for $A'$, the matrix $A'\gamma'^0$ is positive definite by Theorem 5. Equation (\ref{105}) shows that the matrix  $A\gamma ^0$ is then positive definite, too. This completes the proof.\\

\section{Anticommutation relation for the alpha matrices}\label{anticom-alpha}

{\bf Theorem 7.} {\it Let the matrices $\gamma ^\mu$ obey the general anticommutation relation (\ref{Clifford}). In order that the matrices $\alpha  ^\mu$ given by Eq. (\ref{alpha}) satisfy an anticommutation of the form 
\be \label{Clifford-h}
\alpha  ^\mu \alpha  ^\nu + \alpha  ^\nu \alpha  ^\mu = 2h ^{\mu \nu}\,{\bf 1}_4, \quad \mu ,\nu \in \{0,...,3\}, 
\ee
it is necessary and sufficient that the components $g^{0 j}\ (j=1,2,3)$ of the metric be zero. The components $h ^{\mu \nu}$ are then given by}
\be \label{metric-h}
h ^{00}\equiv 1/g ^{00}, \quad h^{0 j}\equiv h^{j 0}\equiv 0\ (j=1,2,3), \quad h^{jk}\equiv  -g^{jk}/g ^{00} \ (j,k\in \{1,2,3\}).
\ee

{\it Proof.} Consider the matrices $\alpha'^0 \equiv g^{00} \alpha^0$ and $\alpha'^j \equiv  g^{00} \alpha^j$, where as defined in (\ref{alpha}), $\alpha ^0 \equiv \gamma ^0/g^{00}, \quad \alpha ^j \equiv \gamma ^0\gamma ^j/g^{00}$. Using (\ref{Clifford}), we find
\bea
\alpha'^0 \alpha'^0 + \alpha'^0 \alpha'^0 & = & 2g^{00} {\bf 1}_4,\label{EA1}\\
\alpha'^0 \alpha'^j + \alpha'^j \alpha'^0 & = & 2g^{0j}\gamma ^0,\label{EA2}\\
\alpha'^j \alpha'^k + \alpha'^k \alpha'^j & = & \left( -\gamma ^j\gamma ^0+2g^{0j}{\bf 1}_4\right)\gamma ^0\gamma ^k+\left( -\gamma ^k\gamma ^0+2g^{0k}{\bf 1}_4\right)\gamma ^0\gamma ^j \nonumber\\
& = & -2g^{jk}g^{00} {\bf 1}_4 + 2\left(g^{0j}\gamma ^0\gamma^k+ g^{0k}\gamma ^0\gamma ^j\right).\label{EA3}
\eea
Setting $g^{0j}=0 \quad (j=1,2,3)$, the result follows then immediately from the definition (\ref{alpha}). Q.E.D.\\

Note that for admissible coordinates (\ref{metric-h}) shows that the quadratic form  $h ^{\mu \nu}dx^\mu dx^\nu $ has Euclidean signature.  (Note that $h ^{\mu \nu}$ is not a tensor.)\\

Theorem 7 restricts the consideration to special affine coordinates such that  $g^{0 j}=0\ (j=1,2,3)$, called Gaussian affine coordinates.  Note that Gaussian coordinates exist in a neighborhood of every event in a Riemannian space-time, but are not necessarily the coordinates of choice for rotating observers \cite{A41}.  Theorem 6 of Appendix \ref{uniqueness-affine} allows us to consider more general coordinates that include rotating coordinate systems.

\bigskip

\end{document}